\documentclass[%
 reprint,
 amsmath,amssymb,
 aps,superscriptaddress
]{revtex4-1}

\usepackage{graphicx}
\usepackage{dcolumn}
\usepackage{bm}
\usepackage{hyperref}
\usepackage{float}
\usepackage{newfloat}
\DeclareFloatingEnvironment[name={FIG. S}]{suppfigure}

\hypersetup{
    unicode=false,
    pdftoolbar=true,
    pdfmenubar=true,
    pdffitwindow=false,
    pdfstartview={FitH},
    pdftitle={Cavity-assisted manipulation of freely rotating silicon nanorods in high vacuum},
    pdfauthor={Stefan Kuhn},
    pdfnewwindow=true,
    colorlinks=true,
    linkcolor=red,
    citecolor=blue,
    filecolor=magenta,
    urlcolor=cyan   
}

\begin{document}

\preprint{APS/123-QED}

\title{Cavity-assisted manipulation of freely rotating silicon nanorods in high vacuum}

\author{Stefan Kuhn}
\email{stefan.kuhn@univie.ac.at}
\affiliation{University of Vienna, Faculty of Physics, VCQ, Boltzmanngasse 5, 1090 Vienna, Austria}
\author{Peter Asenbaum}
\affiliation{University of Vienna, Faculty of Physics, VCQ, Boltzmanngasse 5, 1090 Vienna, Austria}
\author{Alon Kosloff}
\affiliation{School of Chemistry,Tel-Aviv University, Ramat-Aviv 69978, Israel}
\author{Michele Sclafani}
\affiliation{University of Vienna, Faculty of Physics, VCQ, Boltzmanngasse 5, 1090 Vienna, Austria}
\altaffiliation[]{Current address: Institut de Ciencies Fotoniques, 08860 Castelldefels, Barcelona, Spain}
\author{Benjamin A. Stickler}
\affiliation{University of Duisburg-Essen, Lotharstr. 1, 47048 Duisburg, Germany}
\author{Stefan Nimmrichter}
\affiliation{University of Duisburg-Essen, Lotharstr. 1, 47048 Duisburg, Germany}
\author{Klaus Hornberger}
\affiliation{University of Duisburg-Essen, Lotharstr. 1, 47048 Duisburg, Germany}
\author{Ori Cheshnovsky}
\affiliation{School of Chemistry,Tel-Aviv University, Ramat-Aviv 69978, Israel}
\author{Fernando Patolsky}
\affiliation{School of Chemistry,Tel-Aviv University, Ramat-Aviv 69978, Israel}
\author{Markus Arndt}
\affiliation{University of Vienna, Faculty of Physics, VCQ, Boltzmanngasse 5, 1090 Vienna, Austria}


\begin{abstract}
Optical control of nanoscale objects has recently developed into a thriving field of research with far-reaching promises for precision measurements, fundamental quantum physics and studies on single-particle thermodynamics. Here, we demonstrate the optical manipulation of silicon nanorods in high vacuum. Initially, we sculpture these particles into a silicon substrate with a tailored geometry to facilitate their launch into high vacuum by laser-induced mechanical cleavage. We manipulate and trace their center-of-mass and rotational motion through the interaction with an intense intra-cavity field. Our experiments show optical forces on nanorotors three times stronger than on silicon nanospheres of the same mass. The optical torque experienced by the spinning rods will enable cooling of the rotational motion and torsional opto-mechanics in a dissipation-free environment.
\end{abstract}

\maketitle

Nanoparticles often exhibit unique optical, mechanical, or electro-magnetic properties because of quantum effects in confined geometries and low dimensions~\cite{Krahne2013}. 
Complementary to that, our present study is part of a long-term effort to control the quantum properties of the objects' motion \cite{Hornberger2012,Arndt2014,Bateman2014}. 
First experiments demonstrating de Broglie wave optics with macromolecules~\cite{Arndt1999} were triggered by the question whether the superposition principle of quantum mechanics holds on all scales.
They have led to the observation of quantum interference with masses beyond 10,000\,amu~\cite{Eibenberger2013}. 
An even higher mass regime, which might give insight to the quantum-classical transition, can be reached with novel coherent manipulation schemes~\cite{Haslinger2013}. 
Models of a spontaneous localization of the wave function~\cite{Bassi2013}, and non-standard effects of gravity~\cite{Diosi1987,Penrose1996}, will become relevant for delocalized particles in the mass range of $10^5-10^8$\,amu, and above $10^{10}$\,amu, respectively.
Tests of such models will require neutral, size- and shape-selected, cold, and slow nanoparticles that are mechanically isolated from their environment. This has motivated new experiments to launch and cool dielectric nanospheres in optical tweezers~\cite{Li2011,Gieseler2012,Kiesel2013}, ion traps~\cite{Millen2015}, and in free-flight~\cite{Asenbaum2013}. 

Here we extend this research to rod-shaped dielectrics of tailored geometry and anisotropic polarizability. 
Once their rotational motion can be controlled sufficiently well, these nanorods may be suitable for realizing torsional optomechanics~\cite{Shi2013,Yin2013,Mueller2015}. 
Earlier experiments have shown that optomechanical torques can be exerted on nanorods in solution using the polarization or orbital angular momentum of a light field~\cite{Bonin2002,Paterson2001,Jones2009,Tong2010}. The manipulation of nanoparticles in a dissipation-free environment, however, has remained challenging~\cite{Marago2013}.
The coupling between rotational and motional degrees of freedom was recently demonstrated with optically trapped birefringent microspheres in a low pressure environment~\cite{Arita2013}.

\begin{figure}[]%
\centering
\includegraphics[width=0.45\textwidth]{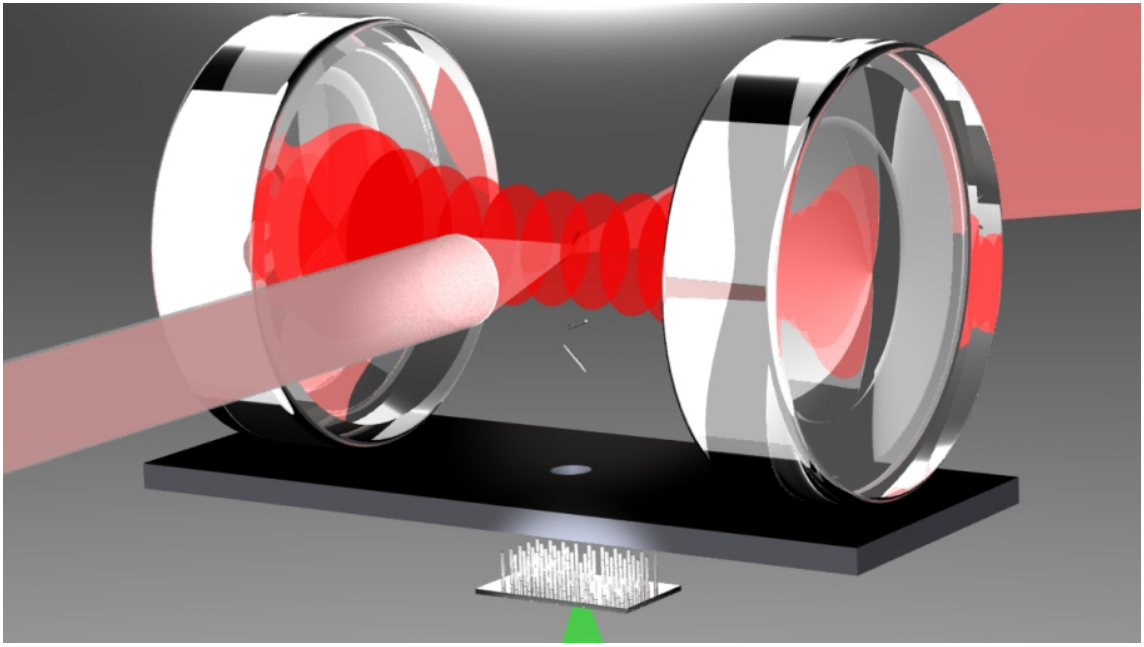}%
\caption{{\bf Launch and detection of size- and shape-controlled silicon nanorotors.} An array of nanorods is carved into a silicon wafer. Focused laser pulses ($1-3\,$mJ pulse energy, 6\,ns duration and $532\,$nm wavelength) hitting the rear side of the substrate cause stress-induced breaking of the nanorods at predefined trenches on the front side and launch them into the high vacuum. Individual, freely moving, and rotating rods need to pass a 500$\,\mu$m diameter and 4\,mm long aperture before entering a strongly pumped high-finesse infrared cavity ($\lambda = 1560$\,nm, 400\,W intra-cavity power, 330,000 finesse, $w_0 = 65\,\mu$m waist). Their velocities and rotation rates are deduced from the scattered light, which is collected by a multi-mode fiber placed perpendicular to the cavity axis and the direction of field polarisation.}
\label{fig:Setup}
\end{figure}

\begin{figure*}[]%
\centering
\includegraphics[width=\textwidth]{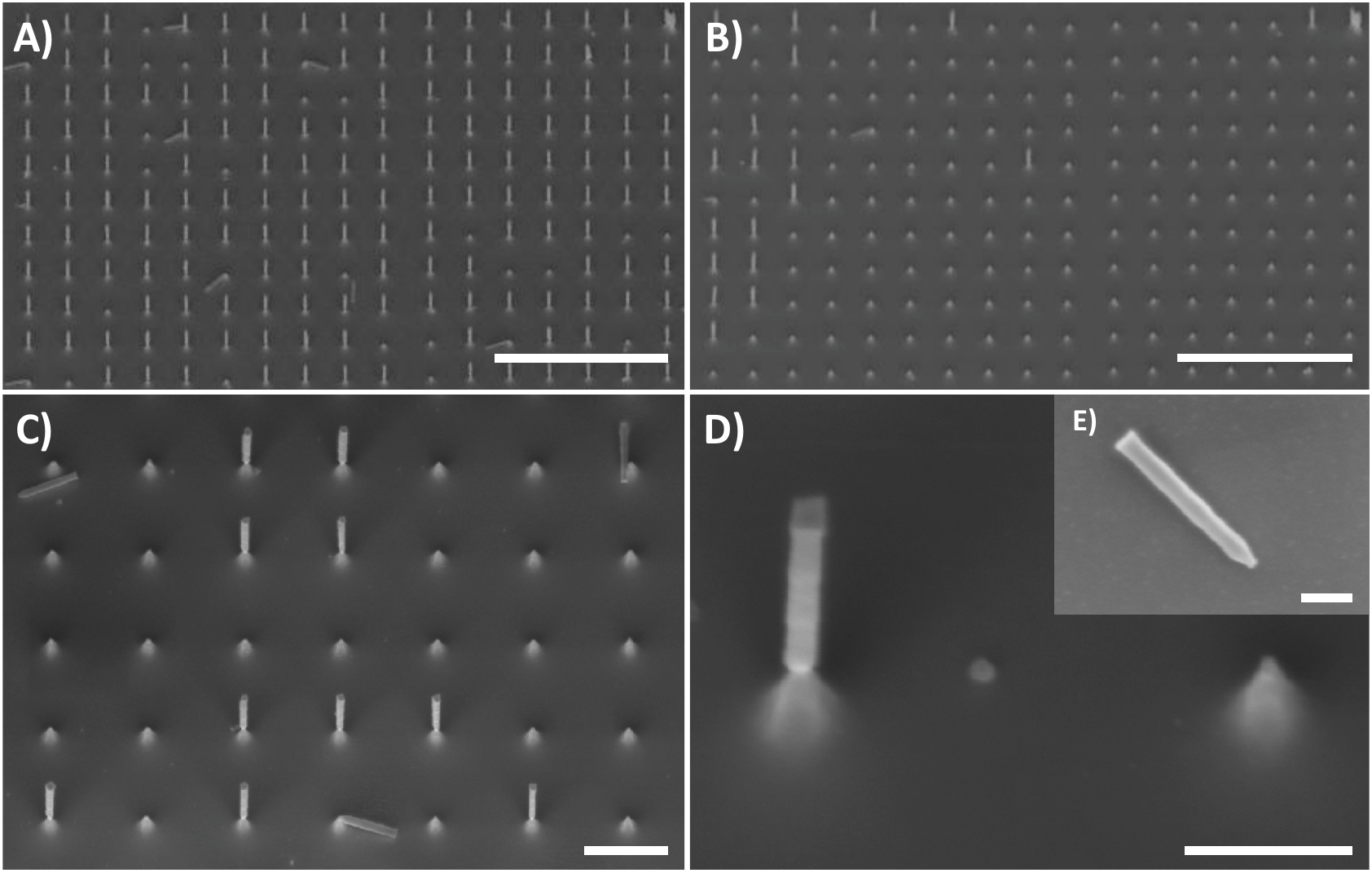}%
\caption{{\bf SEM micrographs of the nanosculptured silicon rods}. 
Array of nanorods with underetched kerfs (A) before and (B) after LITHMOS desorption. C \& D) The kerfs define the break points where the rods crack. The actual rod length appears shortened under the oblique viewing angle. Nanorods with well-defined geometries can be launched, sent through the cavity and collected on a sample plate (E). Scale bars: (A \& B) $5\,\mu$m, (C) $1\,\mu$m, (D) 400\,nm and (E) 200\,nm.}%
\label{fig:SEM}%
\end{figure*}

We have prepared periodic arrays consisting of more than a million silicon nanorods per mm$^2$ by dry-etching crystalline silicon wafers (see Appendix\ref{sec:AppA}). A scanning electron microscopy image of such an array is displayed in Figure \ref{fig:SEM}A. The individual nanorods exhibit a length of $795 \pm 17$\,nm and a diameter of $108 \pm 16$\,nm corresponding to a mass of $(1.0 \pm 0.3) \times 10^{10}$\,amu. The sample was positioned in a chamber evacuated to $10^{-6}$\,Pa underneath an optical cavity (see Figure \ref{fig:Setup}). 
The backside of the sample was locally heated by a pulsed laser beam focused to $100\,\mu$m, which desorbs the rods by laser-induced thermomechanical stress (LITHMOS)\cite{Asenbaum2013}. Small kerfs etched into the nanorod base define the desired break-off conditions (see Figure \ref{fig:SEM}C \& D; details in Appendix\ref{sec:AppA}). In Figure \ref{fig:SEM}A and B we show an electron micrograph of a sample spot before and after the LITHMOS pulses. It demonstrates that the rods can be reproducibly broken off at the tailored constrictions. Figure \ref{fig:SEM}E depicts a close-up of the etched conical tip of a rod after launch and recapture.

The standing light wave field of a high-finesse cavity allows us to track the translational and rotational motion of the particles. It is optically pumped by a linearly polarized, distributed-feedback laser locked close to the cavity resonance.
At the laser wavelength of 1560\,nm silicon exhibits a high relative permittivity and minimal absorption. In a homogeneous field, the polarisability assumes a maximum value of $\alpha_{\|} / 4 \pi \varepsilon_0 = 6.4 \times10^9$\AA$^3$ and a minimum value of $\alpha_{\bot}/4 \pi \varepsilon_0 = 9.8 \times 10^8$\AA$^3$ when the rods are oriented parallel and perpendicular to the field, respectively~\cite{Hulst1957}. Even for rotating rods, the polarizability averaged over all possible rotation axes, $\left(\frac{1}{3} \alpha_{\|} + \frac{2}{3} \alpha_{\bot}\right)/4 \pi \varepsilon_0 = 2.8 \times10^9$\AA$^3$, is larger than for a silicon nanosphere of the same mass, $\alpha_{\rm sph}/4 \pi \varepsilon_0 = 1.4 \times10^9$\AA$^3$.

\begin{figure*} []
\centering
 \includegraphics[width=0.87\textwidth]{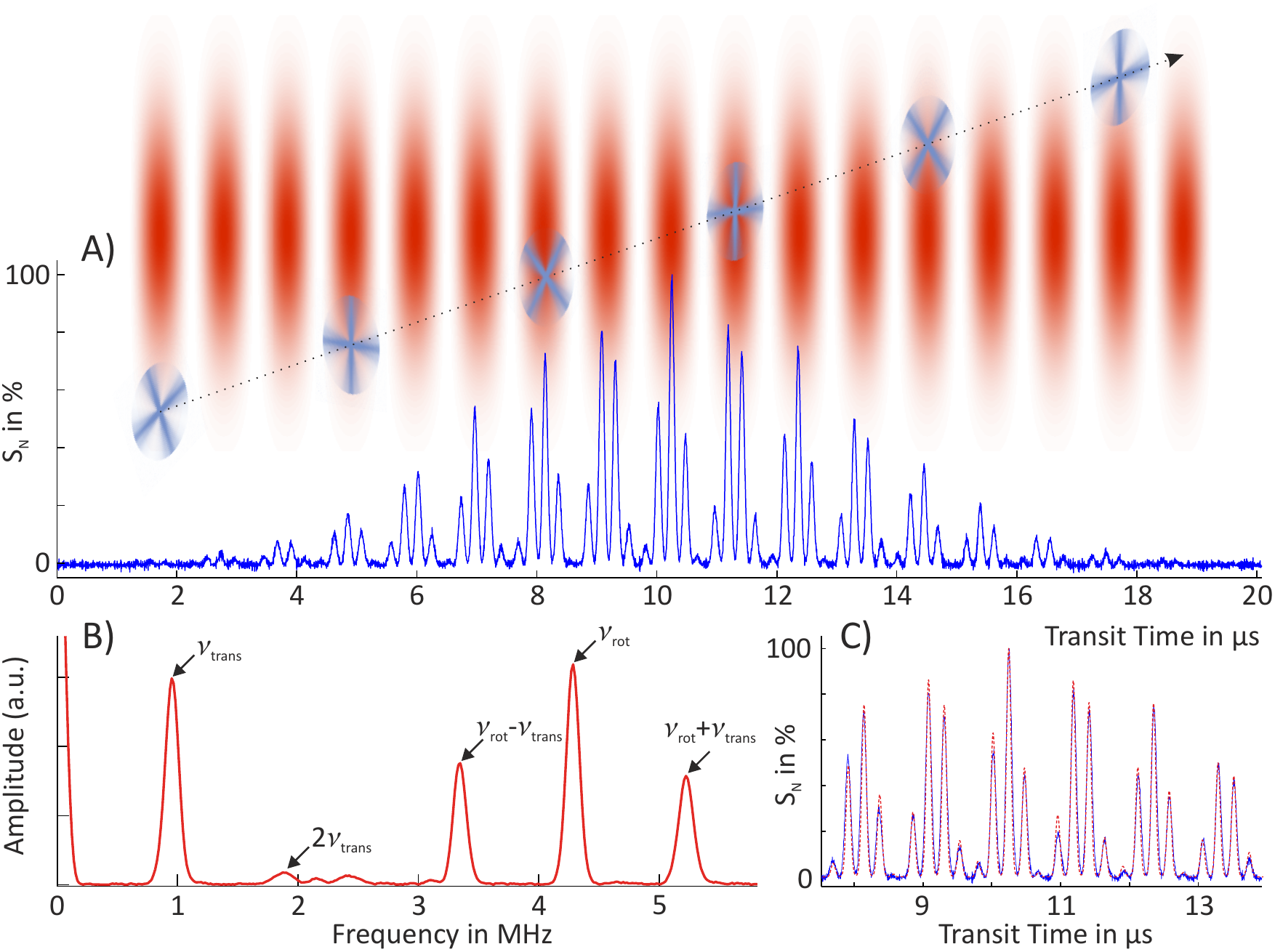}%
 \caption{\textbf{Tracing the nanorotor kinematics.} The motion of every single nanorod can be deduced from the light it scatters while passing the cavity. A) The signal as a function of time conveys three distinct time scales corresponding to (i) the vertical transit through the cavity mode (Gaussian envelope), (ii) the transverse motion across the standing wave (low frequency $\nu_{\rm trans}$), and (iii) the nanorod rotation (high frequency $\nu_{\rm rot}$). B) These frequency contributions can also be identified in the Fourier spectrum of the signal. The particle displayed here exhibits the vertical velocity $v_x=11.5 \pm 0.5 $\,m/s, the on-axis velocity $v_z = 0.77 \pm 0.05$\,m/s and the rotation rate $f_{\rm rot} = 2.15 \pm 0.03$\,MHz. 
The geometric collimation of the incident particle trajectories permits an unambiguous distinction of the translational and rotational motion. C) The measured scattering signal (blue solid line) is well explained by a simple theoretical model (red dashed line), see Appendix\ref{sec:AppB} \& Figure S\ref{fig:S1}.}
 \label{fig:Rotation}%
 \end{figure*}

We can trace each nanorotor using the light it scatters into the direction perpendicular to both the cavity axis and the field polarisation. We collect this light in a 1\,mm multimode fiber placed at a distance of 200\,$\mu$m from the cavity center. The detected intensity depends on the rod's position in the standing wave and also on its orientation (see Appendix\ref{sec:AppB}). 

When a symmetric rotor enters the cavity and moves freely along the cavity axis, we expect a modulation of the scattering signal at two distinct frequencies: One is the translational frequency $\nu_{\rm trans} = 2 v_z/\lambda $ of the particle passing the standing-wave nodes with velocity $v_z$, the second one is twice the rotation frequency, $\nu_{\mathrm{rot}}=2f_{\mathrm{rot}}$.

Figure \ref{fig:Rotation}A displays the normalized scattering intensity of a freely rotating nanorod, $S_N \equiv \left(I_S/I_C\right) / \mathrm{max} \left(I_S/I_C\right)$, where $I_S$ is the measured scattering signal and $I_C$ the simultaneously recorded intra-cavity intensity. Panel C shows that it agrees well with the theoretical expectations for light scattering at dielectric needles, see Appendix\ref{sec:AppB}. We provide a full comparison of the measured signal in panel A and theory in Appendix\ref{sec:AppC} Figure S\ref{fig:S1}. The corresponding Fourier spectrum, depicted in panel B, exhibits the distinct frequency contributions of translation and rotation. 

\begin{figure*} []
\includegraphics[width=0.6\textwidth]{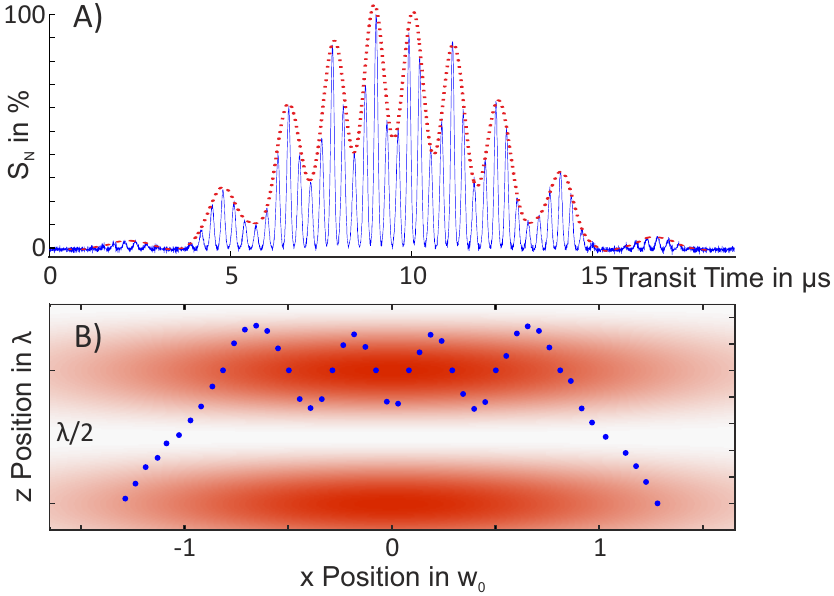}%
\centering
\caption{\textbf{Transverse optical channelling:} A) Scattering signal of a nanorod temporarily captured in an antinode of the standing-wave field. In contrast to Figure \ref{fig:Rotation}, the modulation envelope averaged over the fast rotation (red dotted line) no longer drops to zero between $5-14\,\mu$s. This indicates that the particle is channelled. B) Reconstruction of the center-of-mass trajectory of the rod. Channelling occurs at a trapping frequency around 470\,kHz. A silicon sphere with the same volume would only be trapped at a frequency of 290\,kHz~\cite{Asenbaum2013}.}
\label{fig:Channelling}%
\end{figure*} 

For slow rods we could observe cavity-induced translational channelling, i.e. one-dimensional trapping along an anti-node of the standing light wave. One such case is displayed in Figure \ref{fig:Channelling}, where the scattering signal (panel A) differs significantly from Figure \ref{fig:Rotation}. When averaged over the rotational period of the rod, the scattering signal (red dotted curve) does not drop to zero while the particle is close to the center of the Gaussian beam. During this time the frequency related to the transverse motion of the rod is influenced significantly. In panel B we reconstruct the particle trajectory through the cavity mode from the time evolution of the light scattering curve~\cite{Asenbaum2013}. This is reproduced in a simulation of the rod's motion under the influence of the cavity field (see Appendix\ref{sec:AppC} Figure S\ref{fig:S2}). 

The optical channelling effect benefits from the geometrically enhanced induced dipole moment, due to the strong anisotropy of the rods~\cite{Hulst1957}. For silicon nanorods rotating in the plane perpendicular to the cavity axis the orientation-averaged polarizability in the light field, $\left(\alpha_{\|} + \alpha_{\bot}\right)/2$, is enhanced by a factor of 2.7 in comparison to silicon spheres of the same mass. In Figure \ref{fig:Channelling} we observe an enhancement of the trapping potential by a factor of 2.6.

\begin{figure*} []
\begin{center}
\includegraphics[width=0.9\textwidth]{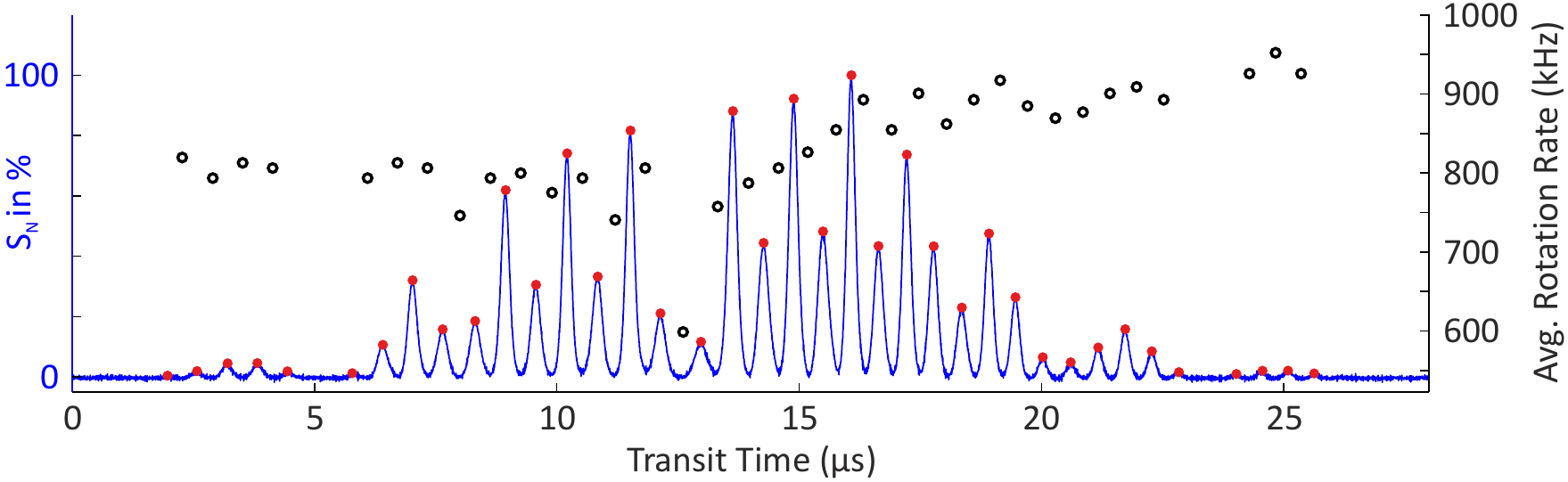}%
\end{center}
\caption{\textbf{Optically induced torque:} Scattering signal (blue curve) of a channelled nanorod, whose average rotation rate (black circles, right scale) is deduced from the separation of two adjacent scattering maxima (red dots), i.e. half a rotation period. The cavity field couples the translational and the rotational motion. We observe that the particle first spins down to 600\,kHz before it speeds up again to beyond 900\,kHz (see also Appendix\ref{sec:AppC} Figure S\ref{fig:S3}).}
\label{fig:RotForce}%
\end{figure*} 
	
In addition to the optical force on the particle's center of mass, the cavity field exerts a torque on the rods, which tends to align them along the field polarisation axis~\cite{Bishop2003}. Figure \ref{fig:RotForce} displays the scattering curve of a slowly rotating nanorod manipulated in both its center-of-mass motion and its rotation. In analogy to Figure \ref{fig:Channelling}, we observe transverse channelling, but the rotation rate (black circles) is now influenced significantly by the optical torque. This indicates that the motional and rotational degrees of freedom exchange energy via the optical potential. The influence of the optical potential on the rotation rate is most pronounced at around 12\,$\mu$s.

In conclusion, we have presented a method to tailor, launch, track, and manipulate high-mass silicon nanorods with well-defined geometry and high aspect ratio. Light scattering inside a high-finesse infrared cavity allows us to follow the translational and rotational motion of the nanorotors in real time. For some of the rods, cavity assisted 1d-trapping and even rotational forces could be demonstrated. Our results are in good agreement with theoretical expectations and show that the rod-like shape enhances the interaction between the particles and the cavity field  significantly, compared to silicon nanospheres of the same mass. Optomechanical trapping and cooling of the center-of-mass motion~\cite{Li2011,Gieseler2012,Kiesel2013,Asenbaum2013,Bateman2014,Millen2015} will benefit from aligning the rods along the axis of polarisation. Recent studies on single particle thermodynamics~\cite{Gieseler2013,Gieseler2014,Millen2014} may be extended to rotating systems. Our results represent a first step towards realizing torsional nano-optomechanics~\cite{Shi2013,Yin2013,Mueller2015} and rotational cooling, which may become applicable to delicate biological nanomaterials such as the similarly shaped tobacco mosaic virus~\cite{Ashkin1987,Romero-Isart2010}. 
 
\onecolumngrid
\section*{Appendix}
\subsection{Nanorod sculpting}
\label{sec:AppA}
The silicon nanorod arrays were fabricated from a single crystalline Si wafer by adapting a previously described dry etching method~\cite{Pevzner2010}. 
The 380\,$\mu$m thick, $<$100$>$ cut and p-doped wafers exhibit a resistivity of 1-10\,$\Omega\,$cm. They were cleaned by sonication, first in acetone then in isopropyl alcohol (IPA), each for 5 min. They were then thoroughly rinsed with deionized water and dried in a stream of N$_2$. The clean Si wafers were spin coated with MMA resist (Copolymer resist EL9, MicroChem) at 4000\,rpm for 60\,s, followed by baking at 180$^{\circ}$C on a hotplate for 2\,min. PMMA resist (Polymer resist A2, MicroChem) was deposited on the MMA layer by spinning at 2000\,rpm for 60\,s, followed by baking at 180$^{\circ}$C on a hotplate for 2\,min.

An array of 250\,nm diameter circular dots with 1\,$\mu$m spacing was written using a Raith 150 ultrahigh-resolution e-beam lithography system (Raith GmbH, Dortmund, Germany). The patterned wafer was developed in MIBK/IPA 1:3 for 1\,min, followed by rinsing with IPA for 20\,s and drying in an N$_2$ stream. A nickel dot array was prepared by e-beam evaporating a 100\,nm thick nickel layer at a base pressure of $10^{-7}$\,Torr with a rate of 1\,\AA/s. Finally, the remaining resist was lifted-off in an Acetone/IPA 1:1 solution, washed with IPA and dried. The nickel nanodot arrays served as masks in the following dry etching. 

Vertical silicon nanowire arrays were fabricated by applying time-multiplexed reactive ion etching in an Inductively Coupled Plasma Deep Reactive Ion Etching machine (ICP-DRIE, PlasmaTherm SLR\,770). In order to form nanopillars with a well defined breaking point, we have implemented a Bosch process, i.e. passivation followed by anisotropic etching. A variation of the ratio between the etching and the passivation times varies the scalloping and leads to different rod diameters. Based on that we set up a three-stage etching process:  Six 'passivation-etching' sequences with a time ratio of 1:1 allowed us to form 700\,nm long nanorod segments. In order to create the breaking points, the ratio of the time windows was logarithmically changed to 0.7:1 during four further steps. Finally, a wider base was formed by changing the interval ratio linearly to 2:1 in six further steps.  The silicon rods were cleaned by removing the nickel caps chemically.

\subsection{Scattering theory}
\label{sec:AppB}
In order to compute the normalized scattered light intensity $S_{N}$, the rods are modelled as thin, homogeneous, dielectric needles of length $L$ and diameter $D$. 
Adopting the scattering theory for dielectric needles~\cite{Schiffer1979} to a standing-wave situation, we find that the light intensity in the direction perpendicular to both the cavity axis ${\bf e}_z$ and the field polarisation axis ${\bf e}_x$ is proportional to 
\begin{align*} 
\frac{I_S}{I_C} \propto k^4 D^4 L^2 \left(\frac{\varepsilon_r -1}{\varepsilon_r +1}\right)^2 \left| {\bf e}_y \times {\bf u}_{\rm int} \right|^2 \left[S_{+}^2 + 2\cos (2k z) S_{+}S_{-} +S_{-}^2 \right] e^{-2(x^2+y^2)/w_0^2},
\end{align*}
with $k=2\pi/\lambda$ the wave number, $w_0$ the cavity waist, and $S_{\pm} = {\rm sinc} \left[ {\bf n}\cdot \left( {\bf e}_z \pm {\bf e}_y \right)kL/2 \right]$. Here, we denote the center-of-mass position of the rod by $(x,y,z)$, and the orientation of the rod is determined by the radial unit vector ${\bf n}$. The internal field points in the direction ${\bf u}_{\rm int} = 2{\bf e}_x + (\varepsilon_r -1) ({\bf n}\cdot{\bf e}_x){\bf n}$. 

\subsection{Simulation}
\label{sec:AppC}
Figures S\ref{fig:S1}, S\ref{fig:S2} and S\ref{fig:S3} display the simulated dynamics of a $L=800\,$nm long and $D=100\,$nm thick silicon rod with mass $M$ and moment of inertia $\Theta = ML^2/12$, which rotates in the plane perpendicular to the cavity $z$-axis. In this case the rod can be treated as a sub-wavelength particle at position $z(t)$, and its orientation with respect to the field polarization $x$-axis is described by $\boldsymbol{n} = \left( \cos \phi (t), \sin \phi(t) , 0\right)$. Given a constant intra-cavity field amplitude $E_0$, the rod's motion is governed by the following classical equations of motion:
\begin{eqnarray}
\ddot{z}(t) &=& - \frac{E^2_0 k}{4 M} \left[\alpha_{\perp} + \left(\alpha_{\|} - \alpha_{\perp} \right) \cos^2 \phi \right] \sin\left(2 k z\right) e^{-2 \frac{\left(v_x t\right)^2}{\omega^{2}_0}} \nonumber \\
\ddot{\phi}(t) &=& - \frac{E^2_0}{4 \Theta} \left(\alpha_{\|} - \alpha_{\perp} \right) \sin\left(2 \phi\right) \cos^2 \left(k z\right) e^{-2 \frac{\left(v_x t\right)^2}{\omega^{2}_0}} \nonumber
\end{eqnarray}
with $\omega_0$ the cavity waist, $v_x$ the vertical velocity of the rod, and $\alpha_{\|,\perp}$ the polarizability components, as given in the main text. 
Particle trajectories that do not pass through the cavity center, but slightly off-axis, can be accounted for by decreasing the field amplitude $E_0$ below the cavity value $\sqrt{4I_C/c\varepsilon_0}$. 
The corresponding scattering intensity is obtained by evaluating the expression given in Appendix~\ref{sec:AppB} along the simulated trajectory. Since the scattering signal is here proportional to $S_{\rm N} \propto \cos^2 \left(kz\right) \exp \left(-2 v_x^2 t^2/\omega^{2}_0\right)$, the center-of-mass trajectory of a freely rotating rod can be reconstructed from the measured scattering signal by averaging over the fast rotation period.
We ensure that we capture the full transit through the cavity mode by carrying out each simulation over the time interval between $\pm 10 \omega_0 / v_x$. 
In order to compare this to the measured data, the scattering signal must be normalized, and the time offset of the simulation must be adjusted, accordingly.

\begin{suppfigure*}[h]%

\includegraphics[width=0.9\textwidth]{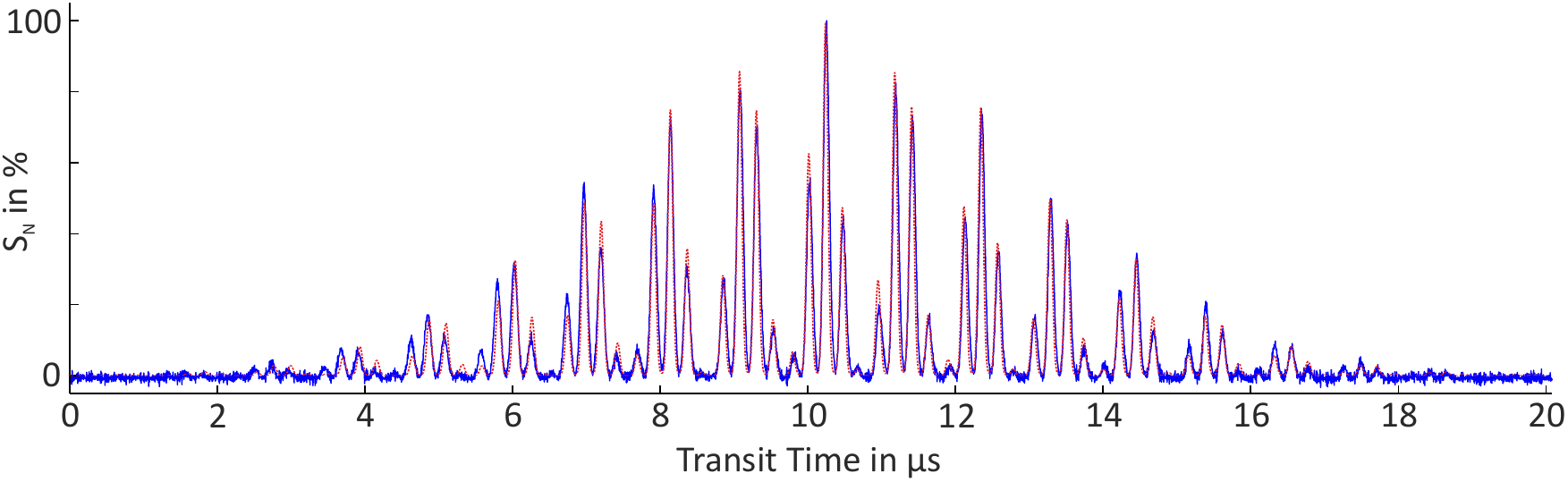}%

\caption{\textbf{Normalized scattering signal of a freely rotating nanorod:} Simulated signal as a function of time (red dashed line) in comparison to experimental data (blue solid line, see Figure \ref{fig:Rotation}). The following parameters and initial values were assumed: $E_0 = 4.15\times 10^6\,$V/m, $v_x = 11.5\,$m/s, $z(0) = -89.85/k$, $\dot{z}(0) = 0.74\,$m/s, $\phi(0) = 0.1\,$rad, $\dot{\phi} (0)/2\pi = 2.14\,$MHz.}

\label{fig:S1}%
\end{suppfigure*}

\begin{suppfigure*}[h]%

\includegraphics[width=0.9\textwidth]{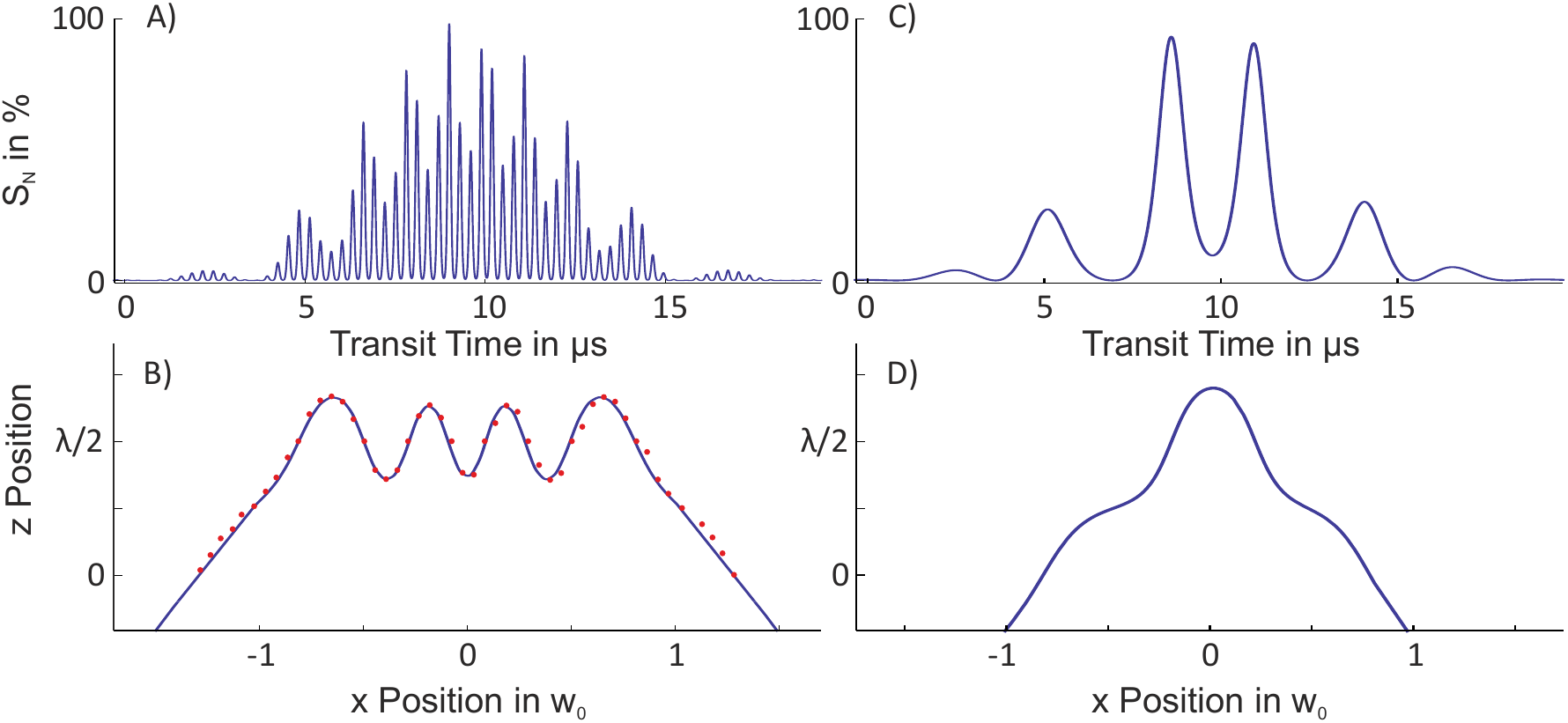}%

\caption{\textbf{Simulation of a 1d-channelled nanorod:} (A) Normalized scattering signal of a 1d-channelled particle. (B) Simulated center-of-mass trajectory along the cavity axis (blue solid line), compared to the trajectory reconstructed from the experimental data (red dots, see Figure \ref{fig:Channelling}). 
The following parameters and initial values were assumed: $E_0 = 8.2\times 10^6\,$V/m, $v_x = 11.3\,$m/s, $z(0) = -91/k$, $\dot{z}(0) = 0.28\,$m/s, $\phi(0) = -0.4\,$rad, $\dot{\phi} (0)/2\pi = 1.685\,$MHz. 
The scattering behaviour (C) and trajectory (D) of a sub-wavelength silicon nanosphere of the same mass is simulated for identical parameters. It illustrates that spherical particles are subjected to weaker optical forces compared to rods.}
\label{fig:S2}%
\end{suppfigure*}  

\begin{suppfigure*}[h]%

\includegraphics[width=0.85\textwidth]{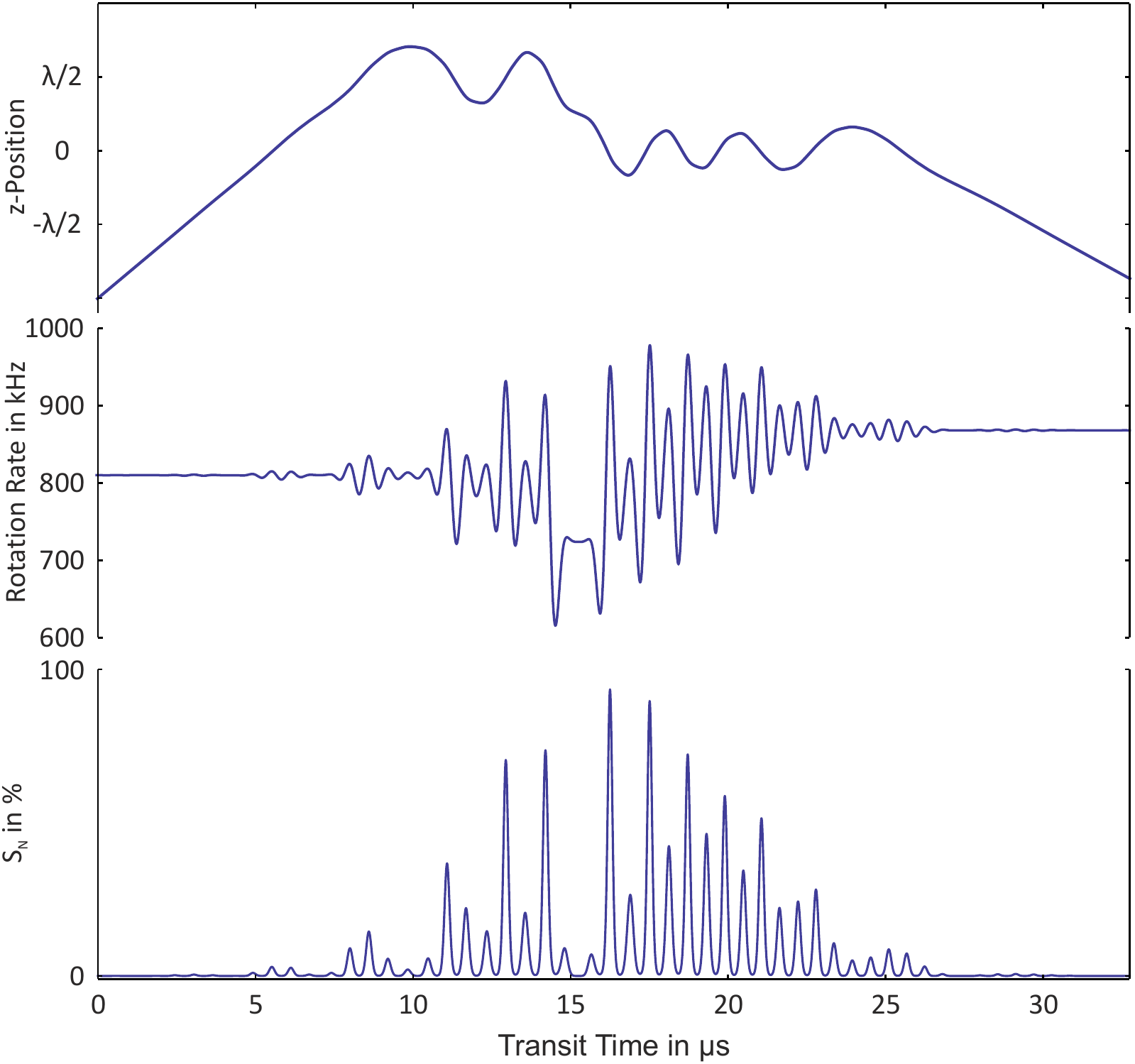}%

\caption{\textbf{Simulation of translational and rotational manipulation of a rod:} This simulation qualitatively resembles the case displayed in Figure \ref{fig:RotForce} of the main text. The motional and rotational degrees of freedom couple via the optical potential, most dominantly after approximately 15\,$\mu$s. At this point, the channelled rod escapes the trapping potential of an anti-node and falls into the adjacent one, while the rotation rate temporarily slows down. 
The following parameters and initial values were assumed: $E_0 = 8.0\times 10^6\,$V/m, $v_x = 7.94\,$m/s, $z(0) = -99/k$, $\dot{z}(0) = 0.28\,$m/s, $\phi(0) = -0.10098\,$rad, $\dot{\phi} (0)/2\pi = 810\,$kHz. 
After transit, we find a 7\,\% higher rotation rate, whereas the on-axis velocity is reduced by 34\,\%.}
\label{fig:S3}%
\end{suppfigure*}

%

\newpage
\section*{Acknowledgements}
Our work has been supported by the European Commission (304886) as well as by the Austrian Science Fund (FWF): W1210-3 and P27297. We acknowledge support by S. Puchegger and the faculty center for nanostructure research at the University of Vienna in imaging the nanorods.
F.P. acknowledges the Legacy Program (Israel Science Foundation) for its support.

\twocolumngrid

\providecommand{\latin}[1]{#1}
\providecommand*\mcitethebibliography{\thebibliography}
\csname @ifundefined\endcsname{endmcitethebibliography}
  {\let\endmcitethebibliography\endthebibliography}{}

\end{document}